# Experimental evidence of a body centered cubic iron at the Earth's core condition

Rostislav Hrubiak[1], Yue Meng[1] and Guoyin Shen[1]*

The crystal structure of iron in the Earth's inner core remains debated. Most recent experiments suggest a hexagonal-close-packed (hcp) phase. In simulations, it has been generally agreed that the hcp-Fe is stable at inner core pressures and relatively low temperatures. At high temperatures, however, several studies suggest a body-centered-cubic (bcc) phase at the inner core condition. We have examined the crystal structure of iron at high pressures over 2 million atmospheres (>200GPa) and at high temperatures over 5000 kelvin in a laser-heated diamond cell using microstructure analysis combined with *in-situ* x-ray diffraction. Experimental evidence shows a bcc-Fe appearing at core pressures and high temperatures, with an hcp-bcc transition line in pressure-temperature space from about 95±2GPa and 2986±79K to at least 222±6GPa and 4192±104K. The trend of the stability field implies a stable bcc-Fe at the Earth's inner core condition, with implications including a strong candidate for explaining the seismic anisotropy of the Earth's inner core.

The Earth's inner core displays directionally dependent elastic properties (i.e. anisotropy) inferred from seismological observations[1]. Aligned hexagonal-close-packed (hcp) phase of iron in the core has been used to interpret the seismic anisotropy [2–4]. However, it has been reported, as well as observed in this study, that the c/a lattice parameter ratio of hcp-Fe becomes more ideal at higher pressures (*P*) and temperatures (*T*) [5–7], effectively lowering the elastic anisotropy. Thus, the aligned hcp-Fe alone is insufficient to account for the observed anisotropy [8–10]. On the other hand, some theoretical simulations have suggested that a body-centered-cubic (bcc) phase of iron may be stable at high *P-T* [11–18]. Aligned bcc-Fe can display a large elastic anisotropy required for explaining the inner core anisotropy [8–10]. However, some other theoretical studies have not replicated the bcc-Fe stability at high *P-T*, e.g. [19,20], or have suggested other possible modifications of Fe [21–23].

Previous experimental studies on the structure of iron at the Earth's core condition have been inconclusive. A study on Fe-Ni alloy [24] showed a bcc structure at high *P-T*. A mixture of hcp-Fe and face-centered-cubic (fcc) iron was reported at 160GPa in samples quenched from high *T* [25], which was interpreted as an indication of bcc-Fe at high *P-T* [26]. Other recent experiments, based on *in-situ* x-ray diffraction (XRD) and x-ray absorption spectroscopy (XAS) measurements in laser heated diamond anvil cell (LH-DAC) [31–36] have suggested hcp-Fe at the core conditions. Data from shockwave compression have suggested (at 75% confidence level) an unknown solid phase other than hcp at 200GPa [27,28]. Later shockwave studies reported no significant density discontinuity associated with the unknown phase [29,30]; however, this doesn't exclude bcc phase since the density difference between bcc and hcp may be small[18]. Whereas, in recent laser driven ramp experiments with XAS [37,38], hcp-Fe was found to be stable to over 500 GPa, although the stability *T* window for bcc-Fe may not be wide enough for observing it in a rapid laser driven ramp/shock. An interesting note is the coordination number of Fe in the laser-ramp XAS study [37], which is consistent with the simulated bcc-Fe at high *P-T* [18].

Here we report experimental evidence of bcc iron at the Earth's core condition using a combined approach of microstructure analysis on T-quenched samples, in-situ XRD at high *P-T*, and spatially resolved scanning XRD microscopy. A variation of this approach has recently been applied for determining the high pressure melting of molybdenum [40]. XRD in LH-DAC experiments can contain signals not only from the center of the heating spot, but also the surrounding regions of relatively low-*T* [39] (Fig. S1). This is usually the case when the heating spot is not sufficiently larger than X-ray beam size and/or the sample is a strong scatter producing strong signal even in X-ray tail area. This contamination of signal from low-T regions may be why the hcp-Fe is so widely reported in literature [31,32,36]. In fact, we also observe the XRD spots of hcp-Fe in all our experiments. Therefore, we emphasize that if we were to rely on *in-situ* high *P-T* XRD measurement alone, without the spatially-resolved microstructure data obtained from *T*-quenched samples, we would likely interpret the *in-situ* data as the evidence of hcp-Fe, just as those from previous LH-DAC experiments [31,32,36]. As demonstrated below, the *T*-quenched microstructure and *in-situ* XRD provide complementary and essential evidence of a new observation of bcc-Fe at high *T*. The spatially resolved XRD microscopy in this study allows us to identify the distribution of different iron phases relative to the heating spot and consequently the temperature distribution, providing a critical step for the observation of bcc-Fe which appears only at high *T*.

### Bcc-Fe signatures in quenched microstructure

In our experiments, Fe samples are embedded between two MgO single-crystal layers with (100)-surface orientations and pressurized in a diamond anvil cell (DAC) [39,41]. The initially compressed Fe shows strong *uniaxial* preferred orientation with the hcp-Fe[002] direction rotated towards the compression direction (Fig. 1a). We heat and quench (i.e. rapidly cool) the Fe samples with a square-modulated laser pulse (with a duration of ~1-5*ms*). We perform *in-situ* XRD measurements before, during, and after each heating pulse. We then subsequently examine the microstructure using scanning XRD microscopy [39] (Figs. S2-4). As expected [3], repeated heating of one location to incrementally increased *T* induces re-crystallization and grain growth, resulting in coarse hcp grains (Fig. 1b). However, the resulting grains still retain the *uniaxial* preferred orientation similar to the initial microstructure, given that hcp-Fe(002) reflections are not visible in the re-crystalized sample. Remarkably, at *P*>95GPa, the Fe samples quenched from *T* above a certain threshold begin to show a microstructure with strong *bi-axial* alignments of hcp-Fe with respect to the (100)-surface oriented MgO in two variations, Fe[001](110)//MgO[100](001) and Fe[001](110)//MgO[010](001), as well as the display of hcp stacking faults appearing as the diffuse streaks as shown in Fig. 1c. The stacking faults involve dislocations in alternating stacking of hexagonal planes and are parallel to the (001) plane, thus generating diffuse streaks along the [001]* reciprocal direction of each of the reflections [42]. Indeed, the streaks observed in the $T$-quenched Fe samples are tangential to the hcp-Fe ($hk$0) rings and intersect with a non-zero angle the ($hkl$) rings with $l\neq 0$ (Fig. 1c) [42]. Strong preferred azimuthal alignment of the streaks, suggests the stacking fault vectors lie preferentially in the plane that is normal to the x-ray beam (parallel to the MgO/Fe surface).

[1]High Pressure Collaborative Access Team (HPCAT), Geophysical Laboratory, Carnegie Institution of Washington, Argonne, Illinois 60439, USA
*Correspondence to: gshen@ciw.edu

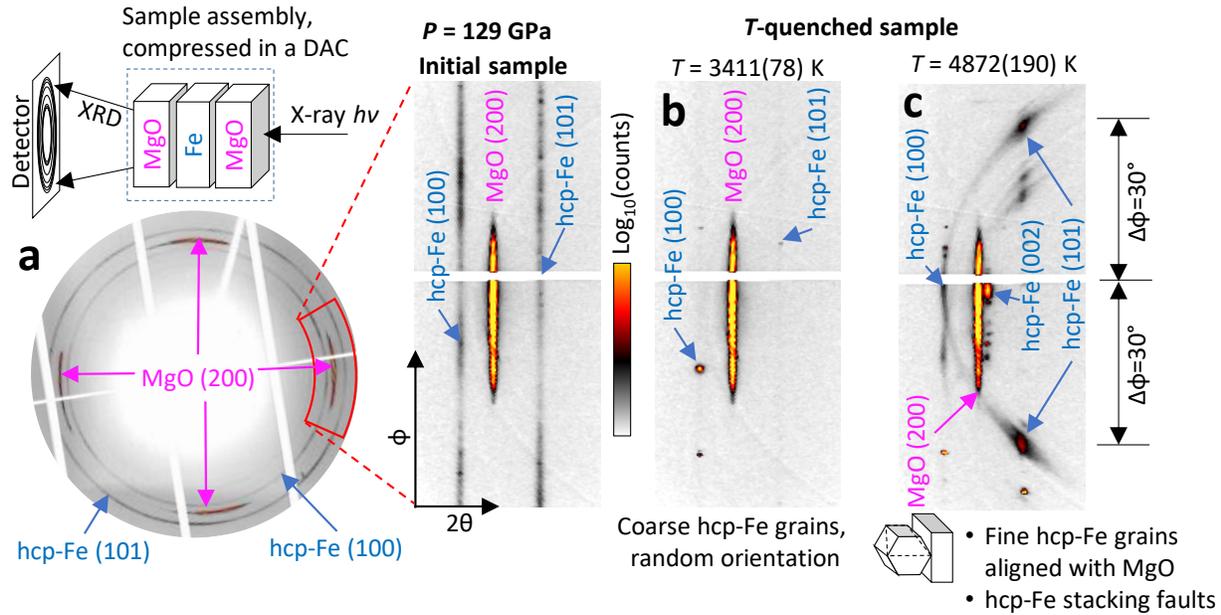

**Figure 1. X-ray diffraction from temperature-quenched samples.** (a) Diffraction geometry and a typical XRD image from compressed MgO/Fe/MgO layers. MgO (200) peaks are aligned to four quadrants in azimuthal (φ) directions. Upon compression, the single crystal MgO layers deform around the Fe sample, resulting in slightly tilted MgO grains with crystallographic orientations close to the initial (100)-surface orientation. (b) XRD after heating to 3411±78K, showing discrete and randomly oriented Bragg reflections, signifying a grain growth with randomly oriented microstructure. (c) XRD after heating to $T$ above a threshold reveals a new, preferred-oriented, microstructure. The hcp-Fe reflections are preferred oriented in azimuthal direction relative to the MgO (200) reflection in each of the four quadrants. XRD diffuse streaks, joining the hcp-Fe (101) and (100) reflections are visible. Region including only one quadrant detail is shown.

We have considered several possibilities to explain the observed *bi-axially* aligned microstructure of hcp-Fe quenched from $T$ above the threshold. Models, based on the epitaxial orientations of hcp-Fe grains directly on the MgO crystal or disordered hcp-Fe at high $T$ [21], are inconsistent with our observations (See supplementary text). An alternative model, consistent with the theoretically predicted bcc-Fe [11–18], is based on the *epitaxial nucleation* of bcc-Fe on MgO crystal and the *growth* of bi-axially oriented bcc-Fe (Fig. S5b) at high $T$ followed by a *topotactic transformation* to bi-axially oriented hcp-Fe upon $T$-quenching (i.e. displacive transformation showing a correlation between crystal orientations of the starting and resulting structures, Fig. S6) [43]. A basic test for this model is whether the orientation of the bcc-Fe phase prior to the topotactic transformation to hcp-Fe (Figs. S6, S7) is consistent with the expected bcc-Fe(001)[110]//MgO(001)[100] epitaxial orientation (Fig. S5b) [44]. Based on the previously studied topotactic orientation relationships between bcc and hcp structures with conjugate planes $(110)_{bcc}//(002)_{hcp}$, $(1\text{-}10)_{bcc}//(0\text{-}10)_{hcp}$ and $(002)_{bcc}//(\text{-}210)_{hcp}$ [45,46], the orientation of the $T$-quenched hcp-Fe crystals with respect to MgO is indeed consistent with and indicative of epitaxial orientation of bcc-Fe on MgO at high $T$ (Figs. S5b, S6, S7).

In addition, the bcc-hcp topotactic transformation is expected to introduce stacking faults due to the inability of atoms to follow all of the entropy states during a rapid quench process [26], in good agreement with our observations. This is also consistent with a previous report [25] showing that stacking faults appear in Fe samples that were statically compressed to 160GPa and quenched from $T$ of 500-1000K below the theoretical melting point. Their observed hcp-Fe stacking faults [25] were interpreted as possible signature of a dynamically stable bcc-Fe at high $P$-$T$ [26].

The heated samples are subsequently decompressed and recovered to ambient pressure. The resulting crystal structure and grain microstructure are examined again using spatially resolved XRD [39]. Strikingly, the release of pressure results in an inverse hcp→bcc topotactic transformation. Bcc-Fe becomes bi-axially aligned to MgO crystals with a preferentially epitaxial orientation, largely returning to the bcc-Fe epitaxial orientation modeled at high $P$-$T$ [44] (Figs. 2, S8). The XRD maps show a direct correlation between the heated locations of the Fe sample and the resulting *bi-axial* alignment even after pressure release. That is, only the heated locations showing the preferred orientation of bcc-Fe at ambient pressure (Fig. 2b), with the remaining region showing no preferred bi-axial alignment (Fig. 2c). The observation of the inverse topotactic transformation clearly supports the model of the direct topotactic transformation (i.e. bcc→hcp upon $T$-quench at high $P$). Furthermore, the location correlation (Fig. 2b) with the heated areas indicates that the epitaxial orientation happens only at high $T$. The lattice parameters of the recovered bcc-Fe samples show identical values to the starting material, within the experimental error, indicating minimum chemical contamination.

Interestingly, there are indications that the epitaxial nucleation and subsequent growth of bcc structured metals on the MgO (100) surface could be a general phenomenon. The same epitaxial orientation, as seen for bcc-Fe on MgO, can be replicated with at least two other metals that are well known to be in the bcc phase at high $P$-$T$ (molybdenum in a previous study[40], and magnesium (Mg) as a part of the current study). We see that Mg becomes epitaxially preferred oriented with respect to the MgO while in the bcc structure at $T>1966$K and $P=35$GPa, and is quenched to preferred oriented $hcp$-Mg at low-$T$ and high-$P$.

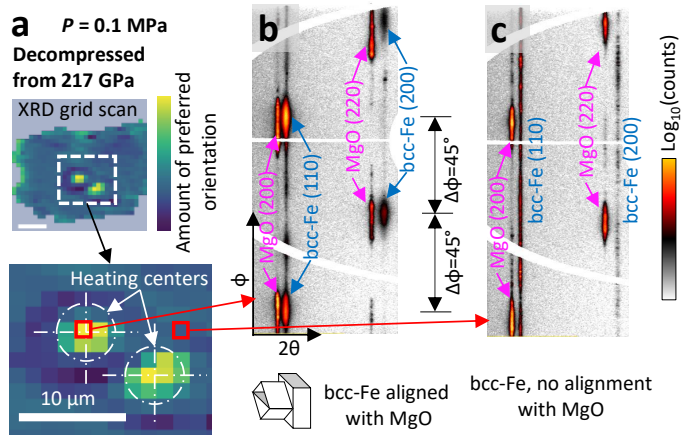

**Figure 3. Spatially-resolved x-ray diffraction maps from a recovered sample.** (**a**) XRD grid scan showing the spatial correlation between the preferred bi-axial alignment in the recovered bcc-Fe and the heating spot locations. (**b**) XRD from a voxel that was heated above the threshold *T*, showing preferentially epitaxial orientation of bcc-Fe to the MgO crystals. (**c**) XRD from a voxel that was heated below the threshold *T*, showing a random microstructure.

### In-situ XRD study

The microstructure evidence of bcc-Fe provides a renewed motivation for us to carefully search for signatures of bcc-Fe using XRD recorded *in-situ* at high *P-T*. XRD images recorded *in-situ* at high *P-T* contain only discrete spot reflections (no continuous Debye rings are visible) due to the large size of grains relative to the x-ray probed volume [39]. Moreover, as we have mentioned, a few XRD spots belonging to hcp-Fe are visible even when the central region of the XRD-probed sample is in the *P-T* field of expected bcc stability. Because the *in-situ* XRD data may be contaminated by hcp-Fe reflections from the surrounding regions at relatively low-*T* [39] (Fig. S1), we have performed scanning XRD microscopy measurements on the *T*-quenched samples, including point-spread-function de-convolution, considering the sizes of the x-ray probe and the x-ray tails [39] (See supplementary text, Figs. S2-S4). Using a correlation between *in-situ* high *P-T* XRD and corresponding *T*-quenched microstructure images, it is unambiguously visible that the coexisting *in-situ* hcp-Fe reflections arise from coarse re-crystallized grains at relatively low-*T* region surrounding the laser-heated area. In view of this, the XRD setup [39] covering a range between 2-7Å$^{-1}$ in *q*-space is then used to search for any other weaker *in-situ* reflections in the recorded XRD images, the expected possible bcc-Fe (110), (200), and (112) reflections, assuming the bcc-Fe density to be close to that of hcp-Fe at given *P-T*. The strongest bcc-Fe reflection, (110), is not visible in any of the *P-T* points examined. Preferred epitaxial orientation of bcc-Fe grains, as described above, is expected to make unlikely the observation of (110) reflection in our diffraction geometry. It should be noted that the absence of the (110) reflections may actually be systemic for the high *P-T* bcc-Fe phase due to disorder in {110} crystal planes at high *P-T*, caused by shear-induced anisotropic plastic flow [47] or avalanche atomic diffusion [48]. However, in a system with disordered {110} planes, some of the orthogonal planes (e.g. {200}, {112}) may still retain order [47]; therefore, reflections from *hkl* planes that are orthogonal to the disordered {110} planes may indeed be observable. Given the *q*-space coverage in this study, the search is then focused on the reflections (200) and (112). The (200) reflections are not visible, possibly due to preferred orientation or degree of disorder in {200} planes of the dynamic bcc-Fe phase. Remarkably, the (112) reflections are indeed visible. As exemplified in Fig. 3, *in-situ* XRD patterns obtained in multiple experiments at 95-222GPa and at high *T* show a few weak but well-defined XRD spots that are consistent with the expected bcc-Fe (112) and cannot be assigned to any reflections of hcp-Fe. The *P-T* conditions at which the bcc-Fe (112) reflection is observed coincide with those corresponding to the observations of the preferred oriented hcp-Fe in *T*-quenched samples (Fig. 4). Moreover, the appearance of the bcc-Fe (112) reflection can be cycled off and on by lowering and raising the temperature in subsequent modulated heating, displaying the reversibility and repeatability of the transition. Even though only one hkl reflection is visible in our experiments, given the fact that the reflections show a lattice spacing predicted for bcc-Fe at corresponding *P-T* conditions, we view it as quite direct evidence. However, we emphasize here that if we were solely relying on the *in-situ* XRD data at high *P-T*, we would be unable to conclude the observation of bcc-Fe at high *P-T* because of the extensive coexistence of hcp-Fe and the weak and sparse nature of XRD spots from bcc-Fe. The microstructural information together with the scanning XRD microscopy of *T*-quenched samples provide essential insight of the existence of bcc-Fe, which is further confirmed by the *in-situ* bcc-Fe XRD at high *P-T*.

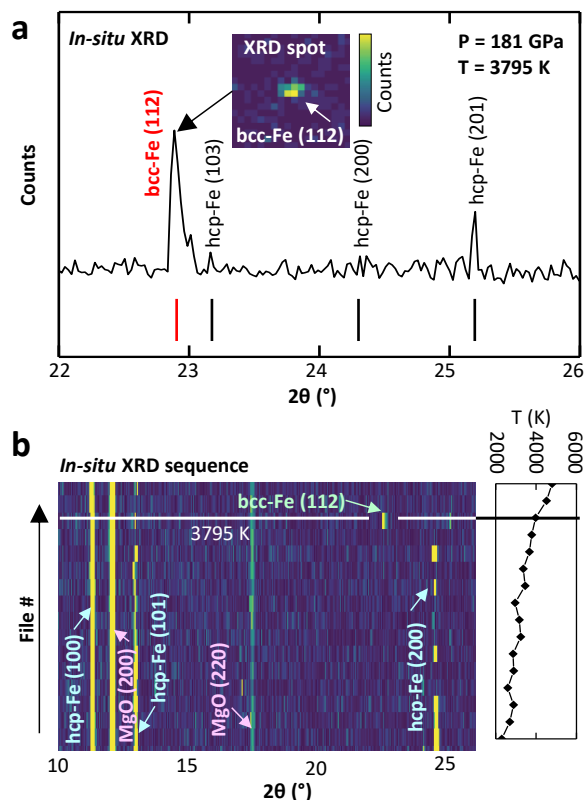

**Figure 2.** *In-situ* x-ray diffraction during a heating pulse. (**a**) Pseudo-powder-XRD pattern recorded from the hot sample showing a bcc-Fe (112) peak, compared to positions of expected hcp-Fe reflections. Insert: an XRD image showing a coarse-grained bcc-Fe (112) reflection. (**b**) Stacked XRD patterns showing the evolution of XRD as a function of *T*, with each stacked row representing a single heating pulse. Bcc-Fe (112) reflections appear exclusively above the threshold *T* of 3795±103K.

### Stability of bcc-Fe

One critical question is whether the observed bcc-Fe is related to the use of MgO in the sample chamber. The oriented (100) MgO surface may play a role in epitaxial nucleation for bcc-Fe given the close 2-D periodicity between the Mg atoms in the MgO surface and the Fe atoms in bcc-Fe. However, metallic layers can only be grown metastably on a substrate up to a few atomic layers [49–52], which would be undetectable by the bulk x-ray probe. Therefore, while the *nucleation* may be heavily related to a substrate (e.g., MgO), the subsequent *growth* into a bulk phase reflects the bcc-stability, independent of the use of MgO. In our experiments, *in-situ* observed bcc-Fe Bragg spots indicate that the bcc-Fe grains are coarse (Fig. 3a insert), not just a few atomic layers. Besides, the appearance of the stacking faults in the *T*-quenched samples (without visible bi-axial alignment) above the transition temperature, was replicated in one experimental run (at a pressure of 103GPa) by replacing MgO with $Al_2O_3$ as encapsulating material. In the case of using $Al_2O_3$, *in-situ* XRD showed an abrupt reduction in the number of hcp-Fe reflections above the threshold *T*, but bcc-Fe (112) reflections were not visible, possibly due to insufficient x-ray counting statistics. We note that the results using $Al_2O_3$ are only preliminary and further work is needed to fully address the influence of MgO and $Al_2O_3$ on bcc-Fe; however, thus far the evidence strongly suggests that bcc-Fe is a thermodynamically stable bulk phase.

The molar volumes, calculated from bcc-Fe (112), are 0-1.4% higher than those of hcp-Fe coexisting in the XRD patterns. Taking the thermal gradient into consideration, the difference in molar volume may be smaller. This volume difference is in general agreement with the predictions for bcc-Fe from molecular dynamics calculations [6,18]. Moreover, previous shockwave studies of Fe have observed a molar volume discontinuity at *T*>4000K-*P*>200GPa with a new solid phase 0.7% less dense than hcp-Fe [27,28], which is consistent with the bcc-Fe phase observed here.

We have performed a total of 20 experiments in a pressure range of 20-222GPa. The hcp-bcc transition is determined by the appearance of bi-axially oriented hcp grains in *T*-quenched samples and by appearance of *in-situ* bcc-Fe (112) XRD reflections at high-*T* in 6 of the experiments above 95GPa. The transition points *T*(*P*) are taken as an average between the lowest *T* with preferentially oriented hcp ($T_{bcc}$) or visible *in-situ* bcc-Fe (112) XRD reflections and the highest *T* without preferentially oriented hcp ($T_{hcp}$) or without visible in-situ bcc-Fe (112) XRD reflections. The *T* uncertainties at each *P* are estimated from the difference between $T_{bcc}$ and $T_{hcp}$, together with the instrumental uncertainty (~100 K). The hcp-bcc transition boundary (Fig. 4) is determined by a linear regression, with weighted uncertainties of *P-T* points (Table S1), giving a functional dependence: $T(K)=2076+9.509P(GPa)$. Furthermore, below 110±3GPa, high-*T in-situ* XRD shows that fcc-Fe is stable at *T* above the bcc-Fe stability field, implying the existence of a triple point of fcc-hcp-bcc. Together with the low-*P* data obtained in this study (20-73GPa), we locate the triple point at 81±13GPa and 2873±212K (3σ certainty level) (See supplementary text, Fig. S9).

The determined hcp-bcc boundary (Fig. 4a) is in good agreement with the solid-solid transition by the shockwave experiment at *P*>200GPa and *T*>4000K [12,27], as well as the predicted hcp-bcc boundary obtained by theory[6]. At *P*>95 GPa, the *P-T* conditions for the hcp-bcc transition are remarkably similar to those associated with previously observed changes in Fe based on visual observation [53,54], *in-situ* XRD [5,31], or XAS [35,37,38] (Fig. 4a). These changes have been attributed to melting [5,35,37,38,53,54], or in one case [31] to "fast" re-crystallization. In light of the new evidence presented here, the hcp-bcc transition may provide an alternative interpretation for the previously reported observations.

**Oriented bcc-Fe in the Earth?**

Previous theoretical studies have suggested [8–10] that bcc-Fe is a candidate for the observed seismic anisotropy. Because the Earth's inner core temperature [31] is likely above the extrapolated hcp-bcc transition line (Fig. 4b), our results imply that Fe adopts the bcc structure under the Earth's inner core condition. However, the previously suggested possible coexistence of both bcc and hcp phases in the inner core [10] remains unclear, given the uncertainties in temperature of the Earth's inner core and the current uncertainties in the hcp-bcc transition line extrapolated to the inner core conditions (Fig. 4b). Finally, a necessary prerequisite for the anisotropy models [8–10] is a mechanism for preferred orientation of the Fe phase under external driving forces, e.g. varying gravitational or Maxwell stress [2]. Our experimental results provide information on how bcc-Fe grains can be nucleated and grown with preferred orientations, relative to its surrounding driving forces. Further studies using our new approach are warranted to understand the effects of alloying elements (i.e. nickel and light elements) on the stability of the bcc-Fe, and vice-versa, the effect of bcc-Fe on element partitioning in the Earth's inner core.

**Methods**

Methods, and any associated references, are available online in the Supplementary Information.

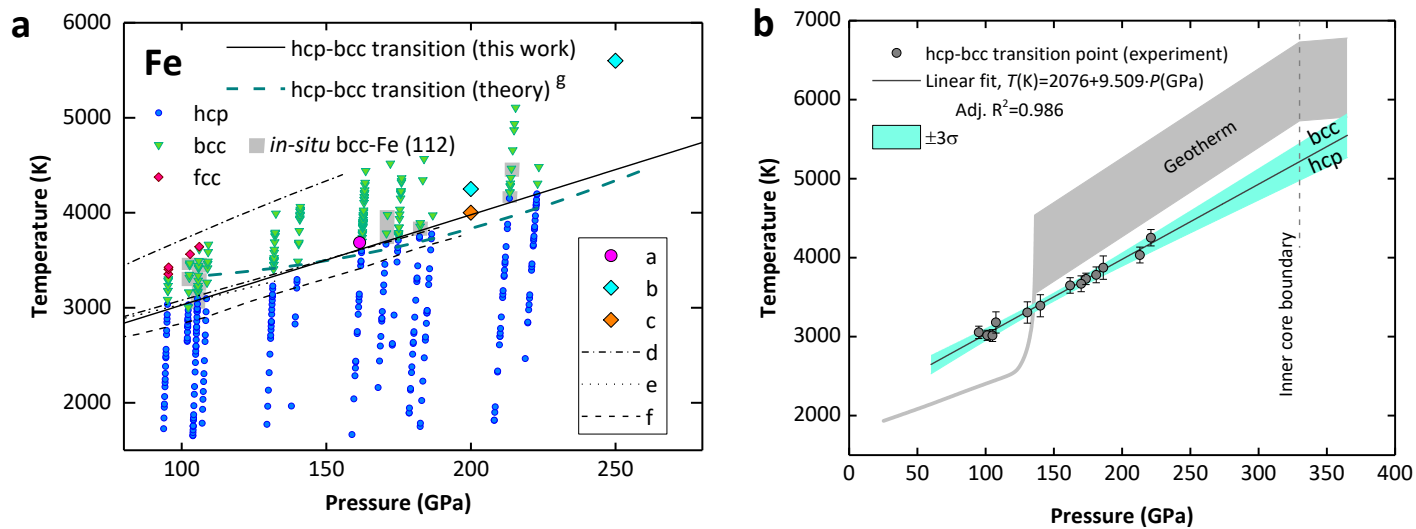

**Figure 4.** The hcp-bcc phase transition and its extrapolation to the Earth's inner core condition. (**a**) Pressure-temperature conditions of the bcc-hcp transition. a: the *P-T* condition where stacking faults are observed in *T*-quenched Fe [25]; b: data points based on shockwave density discontinuities [27]; c: data from shockwave density discontinuity [updated from [55]]; d: Fe melting from XRD [31]; e: upper curve – Fe melting; lower curve – "fast" re-crystallization from XRD [5]; f: Fe-melting from XAS [35]. (**b**) Extrapolation of the hcp-bcc transition with a linear fit to the inner core condition, showing that Fe adopts bcc structure in the Earth's inner core [31,56].

## Acknowledgements

We thank Curtis Kenney-Benson, Stanislav Sinogeikin, Erik Rod, Jesse Smith, and Arunkumar Bommannavar for technical support. We thank Peter Lazor for useful discussions. We also thank anonymous reviewers for critical comments. GS acknowledge the support of DOE-BES, Division of Materials Sciences and Engineering, under Award No. DE-FG02-99ER45775. HPCAT operations are supported by DOE-NNSA under Award No. DE-NA0001974, with partial instrumentation funding by NSF. The Advanced Photon Source is a U.S. Department of Energy (DOE) Office of Science User Facility operated for the DOE Office of Science by Argonne National Laboratory under Contract No. DE-AC02-06CH11357

## Author contributions

R.H. and G.S. designed the study; R.H. and Y.M performed experiments; R.H. analyzed data; R.H. and G.S. interpreted data; R.H. wrote the paper, with contributions from all authors.

## Additional information

Supplementary information is available in the online version of the paper. Correspondence and requests for materials should be addressed to G.S.

## Competing financial interests

The authors declare no competing financial interests.

SUPPLEMENTARY INFORMATION

# Experimental evidence of a body centered cubic iron at the Earth's core condition

R. Hrubiak, Y. Meng and G. Shen

Correspondence to: gshen@ciw.edu

**This PDF file includes:**

Materials and Methods
Supplementary Text
Figs. S1 to S9
Table S1

## Materials and Methods

## Samples and materials

Iron (Fe) foil, Puratronic, 99.995%, purchased from Alfa Aesar, is manually thinned and cut into pieces of required dimensions using steel and tungsten micro-tools. MgO [100μm, (100)-plane orientation] and $Al_2O_3$ [500μm, (11-20) plane orientation] single crystal plates are obtained from MTI Corporation and polished down to a thickness of ~5-10μm. Rhenium gaskets, as well as MgO, $Al_2O_3$ parts for the sample encapsulation assembly, are prepared using a laser micro-machining system at HPCAT (1). Magnesium (Mg) powder, 99.8%, -325 mesh from Alfa Aesar was pressed into a foil and further processed analogously to Fe.

Micro machined encapsulation layers (MgO or $Al_2O_3$) are then heated in an air furnace above 1,100°C in order to remove impurities, particularly $Mg(OH)_2$ in case of MgO. Sample and insulation layers are stacked inside a cylindrical (in some cases rectangular) hole in a pre-indented rhenium gasket (1). The entire diamond anvil cell (DAC) assembly is heated to above 110°C, in a glovebox (Innovative Technology PL-2GB-IL-GP1) with less than 0.1 parts per million (ppm) $H_2O$, in order to remove any adsorbed water from the sample assembly, immediately prior to sealing the DAC pressure vessel. Symmetric type DAC in combinations with cubic-BN seats and diamonds with culets in range of 70-300μm are used in this study to generate high pressure.

## Laser heating with *in-situ* x-ray diffraction (XRD)

At the various pressure points in this study, the main observations are based on XRD measurements at various stages: prior, during, and after a modulated pulse laser heating. For several experimental runs, the observations also include spatially-resolved XRD after heating on *T*-quenched samples, and on *P*-released samples that are recovered to the ambient pressure. Spatially-resolved XRD is described in subsequent section. The temperature range is between 1,500 and 5,300K with intervals ~10-50K.

Laser heating system

The sample is irradiated from two sides simultaneously with slightly de-focused, co-axial, infrared laser beams (λ=1064nm) (2). The shape of the incident laser-power distribution at the sample surface can be approximated by a Gaussian distribution. The full width at half maximum (FWHM) of the laser power distribution is adjustable and in the range of 10-55 μm in diameter. At each pressure point, a sequence of high temperature states are sustained for 1-5 milliseconds by a heating pulse, concurrent with *in-situ* recording XRD from the hot volume.

Fast *T* measurement

Thermal radiation is measured multiple times (>2) during a given heating pulse (1-5 milliseconds) in order to monitor temperatures during a heating pulse using an EM-ICCD detector (PI-MAX4, Princeton Instruments) (2). In several experiments, total thermal emission from the sample is also monitored with a fast photodiode (μs time resolution) to rule out short-duration temperature spikes associated with the Q-switched laser pre-pulse spike. In this study, the pre-pulse spike is eliminated through a step modulation.

Precise alignment

A microscope imaging system with a gated CMOS camera, as previously described (2), is used to collect *in-situ* 2-dimensional (2D) images of the thermal radiation from the laser-irradiated sample surfaces to ensure laser/sample/XRD alignment. Thermal radiation in the range of 600-800 nm is recorded from a circular areas of 4 μm in diameter, axially aligned with the x-ray focus on both sides of the sample (*2*).

In-situ XRD

*In-situ* XRD is performed in axial geometry using micro-focused synchrotron x-rays, with a beam size at the sample position of 3-5 μm vertical by 4-7 μm horizontal FWHM (*2*). A Pilatus 1M-F detector is used to collect diffraction images. The sample-detector geometry allows to cover a range between 2 to 7 Å$^{-1}$ in *q*-space. For XRD images showing continuous Debye rings, radial integration (binning) is performed on the area diffraction images using the XDI software (*3*). D-spacings of Fe and MgO reflections are calculated by fitting Pseudo-Voight functions to corresponding peaks in the radially integrated diffraction patterns. For coarse-grained samples, *d*-spacings are obtained by fitting a 2-D Gaussian function to Bragg spots.

The thermal pressure is calculated directly by using the observed lattice parameters in conjunction with an hcp-Fe equation of state (EOS) from literature (*4*). This produces pressures during the heating pulses with values 1-7% higher than the corresponding cold pressure.

Effect of thermal gradient on *in-situ* XRD

A laser heating spot size can be affected by multiple experimental factors, such as laser focus size and power distribution, available laser power, thickness and thermal conductivity of insulating layers, and temperature range. In some cases, the hottest volume of the sample (above 90% of maximum temperature level) can become comparable in size to the focused x-ray probe's FWHM (Fig. S1). This leaves a large fraction of the tail of the x-ray probe intersecting with colder volume of the sample, causing contamination of recorded x-ray pattern by materials at low temperatures. In the current experiments at pressures over ~100GPa, the volume of the transformed bcc-Fe at high *P-T* is <5 μm in diameter, surrounded by a colder volume of coarse hcp-Fe grains. These hcp-Fe grains intersect with the x-ray probe tail (Fig. S1). Therefore, the *in-situ* XRD patterns that display bcc-Fe reflections also contain coexisting reflections from hcp-Fe located at *T* below the transition in the surrounding volume.

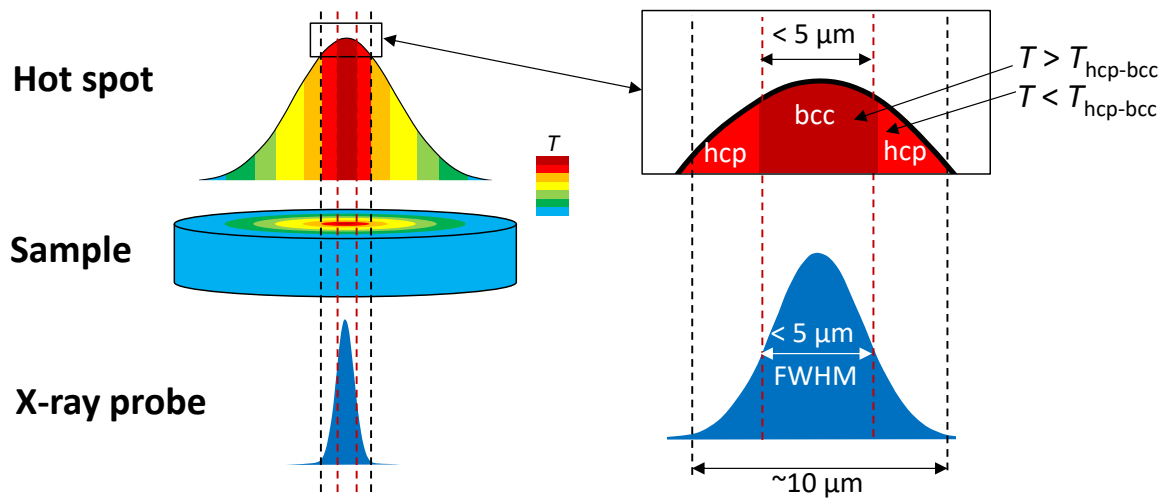

**Fig. S1.** *In-situ* XRD patterns may contain signals from materials at surrounding low temperature region. For an x-ray beam size of ~5 μm in diameter at FWHM, which is much less than a heating spot size of ~15 μm in diameter at FWHM, the bcc-Fe volume is only about 5 μm in diameter surrounded by hcp-Fe at low temperatures, thus displaying the coexistence of bcc-Fe and hcp-Fe in the *in-situ* XRD patterns.

## Scanning XRD microscopy probing microstructure of *T*-quenched Fe samples

XRD measurements are conducted *in-situ* at high *P-T* and on *T*-quenched samples. For some runs, spatially-resolved XRD probing is performed on *T*-quenched samples before and after the hcp-bcc transition. The obtained microstructure maps are used to correlate the hcp-Fe positions with those that appear in *in-situ* XRD at high-*T* with respect to the heating spot. The procedure described below is used to demonstrate that the hcp-Fe Bragg spots coexisting with bcc-Fe in the *in-situ* XRD patterns originate in the low-*T* volume outside the bcc stability field.

Scanning XRD microscopy: spatially-resolved XRD

Unlike a typical high-pressure XRD, which is collected from a single point on a sample using a micro-focused X-ray probe, we obtain spatially-resolved XRD by collecting XRD patterns from a 2-D array of points on a specified rectangular grid over a region of interest (ROI) on the sample, covering a heated volume (Fig. S2A). A step size of 1-2μm is used. Any single XRD pattern (Fig. S2B) in the grid may either show coarse hcp-Fe grains with random orientation or, in cases of samples at *P* > 95 GPa and *T*-quenched from *T* higher than the bcc transition, display XRD texture features corresponding to fine-grained bi-axially oriented hcp microstructure.

XRD intensity from each coarse grain is used to visualize the spatial locations for the corresponding grains (Fig. S2C). In addition, XRD textures are used to map microstructural features by applying chosen ROIs, such as for fine-grained or bi-axially oriented features. These maps, corresponding to spatial distributions of coarse grains (Fig. S2D) and bi-axially aligned microstructure (Fig. S2C), are combined into a composite microstructure map, with each feature assigned a false color, using the XDI software (*3*).

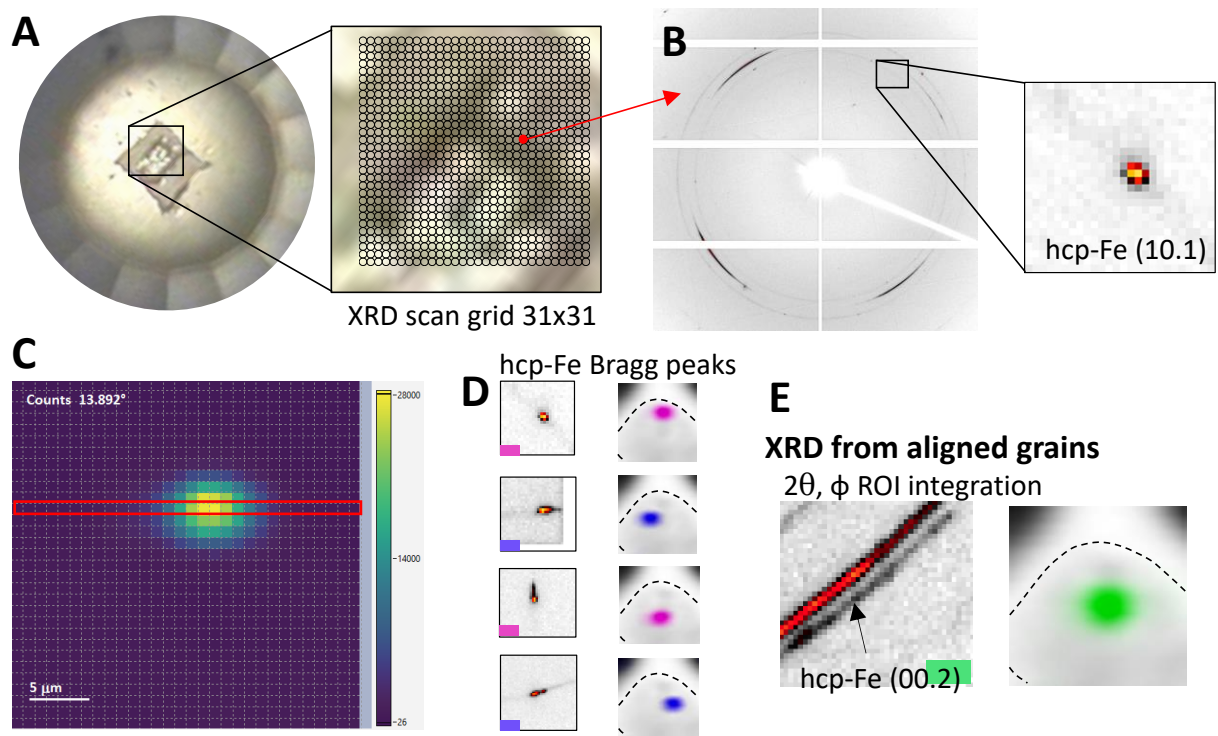

**Fig. S2.** Illustration of spatially resolved 2-D microstructure mapping using XRD. (A) Microscope photograph of a Fe sample compressed inside a DAC, showing an overlaid 2-D scan grid on the right. (B) Typical XRD pattern from a single point on the 2-D scan grid. (C) Intensity plot to visualize the location of one coarse grain with the same XRD orientations. (D) Examples of multiple coarse grains mapped: left column – XRD features of each grain; right column – 2-D map of the grain locations from the corresponding XRD features; colors assigned arbitrarily. (E) Mapping of bi-axially aligned fine-grained microstructure: left – XRD from hcp-Fe (00.2); right – 2-D map of the spatial distribution of the aligned microstructure.

The deconvolution of the x-ray probing size

A simple composite map results in strong overlaps of color-coded features (Fig. S3A). The overlap occurs because the size of the x-ray probe is relatively large compared to the sizes of the microstructural features (Fig. S3B). By applying a deconvolution approach (*5*, *6*), the reconstructed composite map shows much reduced overlap (Fig. S3C). The deconvolution procedure is also carried out by using the XDI software (*3*). The deconvolution procedure is calibrated by reconstructing the spatial position and size of a reference standard gold sphere (2.5 μm in diameter).

Correlating the location of laser heating with the quenched microstructure

Quenched microstructure maps are used to pinpoint the location of individual grains that were subjected to high temperatures. The hcp-Fe coarse grains that appear *in-situ* at high $T$ are quenched to ambient $T$ with their orientations preserved. This allows to trace a single hcp-Fe grain from $T$-quenched XRD to the *in-situ* XRD Bragg spots. As exemplified in Fig. S4, in the XRD pattern collected *in-situ* at high $T$ (in the bcc $P$-$T$ stability range), we see a coexisting hcp-Fe (101) Bragg spot (Fig. S4A). In the $T$-quenched XRD, we again see a Bragg spot appeared in similar orientations likely from the same hcp-Fe grain (Fig. S4B). Using the $T$-quenched microstructure map of the corresponding hcp-Fe grain (Fig. S4C), we obtain the spatial location of this grain relative to the heating spot and to the center of the *in-situ* XRD probe (Fig. S4D). Thus, the hcp-Fe Bragg spots coexisting with the bcc-Fe reflections clearly originate in the low-$T$ volume outside the bcc stability field.

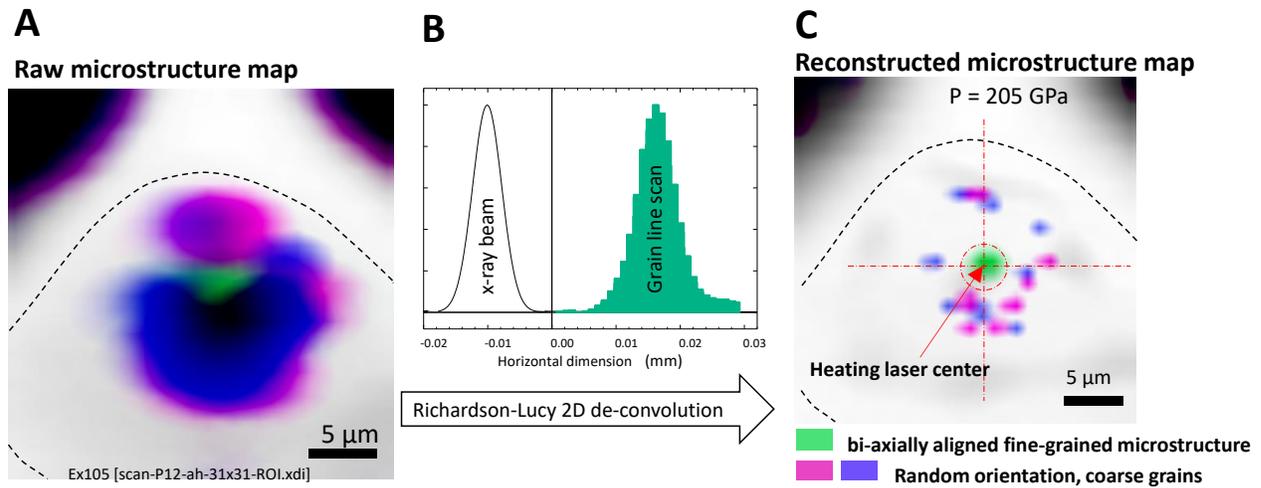

**Fig. S3.** 2-D microstructure map before and after applying the deconvolution method of Richardson-Lucy. (A) A raw composite map showing randomly aligned coarse grains (cyan and magenta) and aligned fine-grains (green). (B) X-ray probe size relative to a line scan of one of the coarse grains. (C) A deconvoluted microstructure map, showing locations of coarse Fe grains relative to the heated spot.

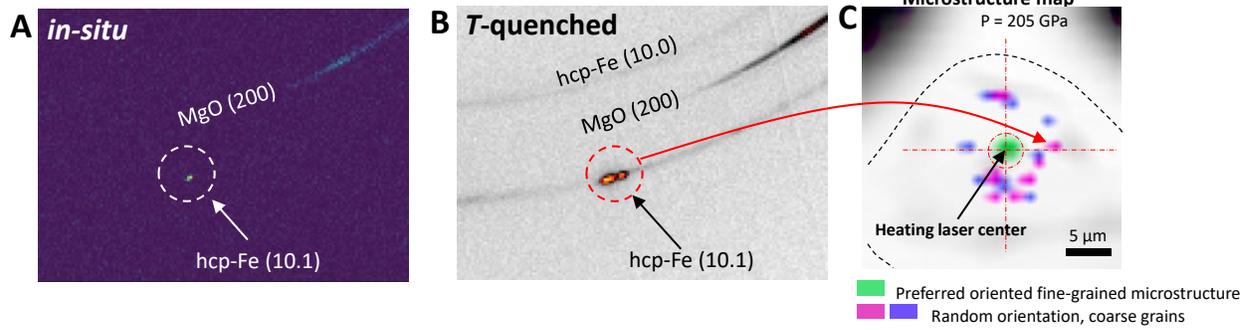

**Fig. S4.** Determining 2-D spatial locations of hcp-Fe grains observed *in-situ* at high *P-T*. (A) Hcp-Fe reflection coexisting in XRD pattern with bcc-Fe. (B) The same orientation as in A is traced to the quenched sample XRD. (C) A microstructure map to visualize the spatial locations of the *in-situ* observed hcp-Fe grains relative to the heated spot.

## Supplementary Text

### Other possible models for the bi-axially aligned hcp-Fe

Several possibilities have been considered to explain the bi-axially aligned microstructure of hcp-Fe quenched from $T$ above the threshold.

One tested model is based on the epitaxial orientations of hcp-Fe grains directly on the MgO crystal at high temperature (Fig. S5A). This model was given some weight at first due to the fact that epitaxial orientations of hcp cobalt (Co) and hcp permalloy (Py: $Ni_{0.8}Fe_{0.2}$, a structural analog to hcp-Fe at ambient pressure) on MgO substrates show essentially the same crystallographic orientations in thin film growth (*7–10*). However, the hcp-Co and hcp-Py depositions on MgO (100) surface are reported to only produce a few stacking faults, localized at the deposition interface (*7–10*). Such an amount of stacking faults would not be detectable by the x-ray probe in this study. Moreover, the hcp-Co and hcp-Py epitaxy defects are expected to be determined by geometrical constraints of already nucleated grains instead of predominant slip systems (*7–10*), which is inconsistent with the experimentally observed preferentially oriented stacking faults as shown in Fig. 1C in the Main text.

Another tested model is related to a suggestion (*11*) that a disordered hcp-Fe structure with stacking faults may be stable at high $P$-$T$. However, the calculations (*11*) show a gradual increase in stacking faults in hcp-Fe beginning at ~500K, which is inconsistent with our experimental data showing that the highly oriented hcp stacking faults appear in quenched samples only after heating to $T$ above a threshold (e.g., >3069K at ~95GPa).

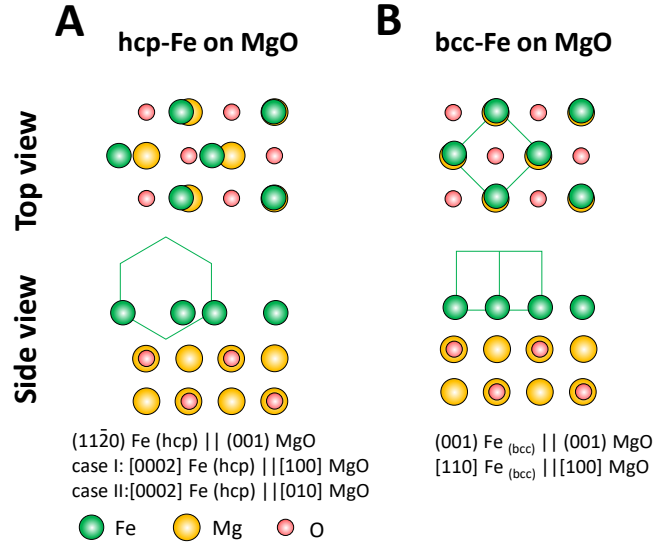

**Fig. S5.** Schematics showing epitaxial plane relationships. (A) Hcp-Fe on MgO epitaxial orientation model. Two possible cases of hcp unit cell orientation, mutually rotated by 90° in the MgO (001) plane. (B) Bcc-Fe epitaxial orientation model. Bcc-Fe and MgO surfaces show a nearly identical 2-D periodicities.

## Topotactic transformation bcc-hcp

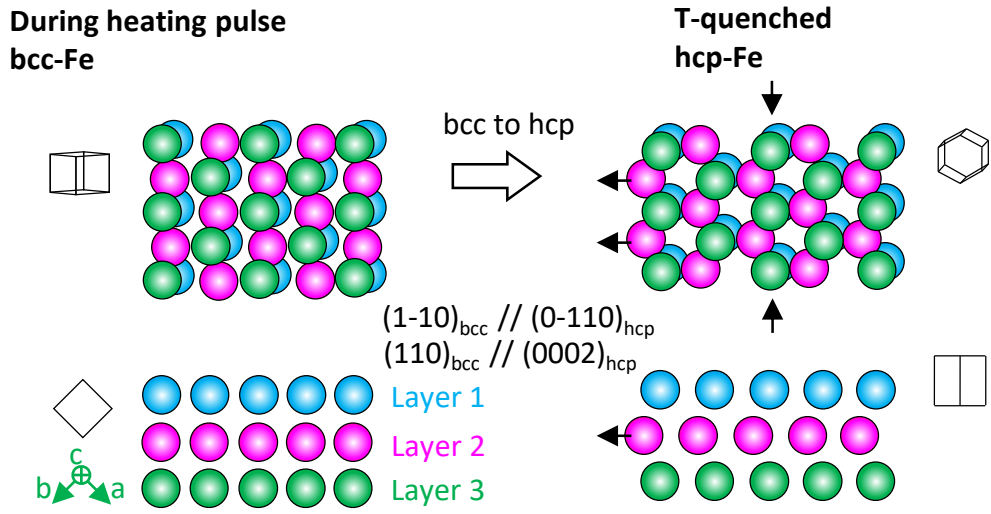

**Fig. S6.** Schematics of a topotactic transformation and relative orientations of bcc-hcp structures.

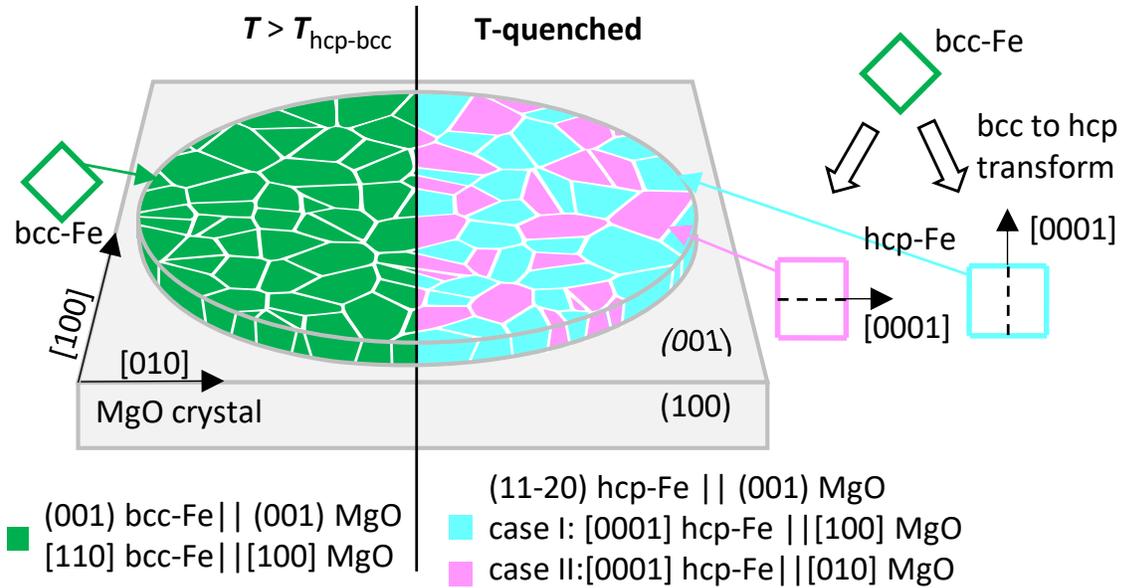

**Fig. S7.** Schematics of the bi-axially oriented bcc-Fe on MgO crystal at high temperature – followed by a topotactic transformation to bi-axially oriented hcp-Fe, upon temperature-quench. The observed orientation in the quenched hcp-Fe are consistent with the established topotactic relations between bcc and hcp structures.

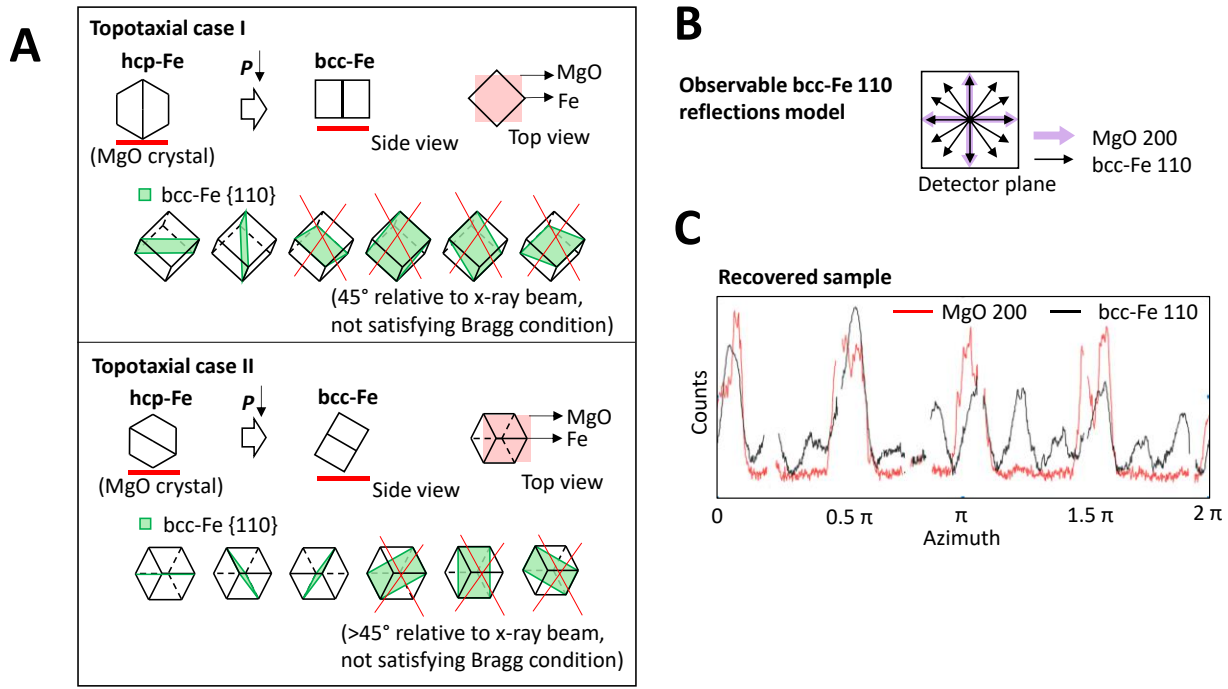

**Fig. S8.** Bcc-Fe orientation based on a topotactic transformation from hcp-Fe upon pressure release. (A) Topotactic case I – bcc-Fe is produced with no rotation on [110] axis with respect to the underlying MgO crystal. From a family of {110} planes, only two are favorably aligned with the XRD setup geometry to satisfy the observable Bragg condition. Topotactic case II – bcc-Fe is produced with 60° rotation on [110] axis with respect to the underlying MgO crystal. From a family of {110} planes, three are favorably aligned with the XRD setup geometry. (B) Schematics showing azimuthal orientations of observable bcc-Fe reflections according to the topotactic model. Maxima are expected in the azimuthal directions coinciding with the MgO (200) reflections as well as at offsets of ~30° multiples relative to MgO (200) reflections. (C) Azimuthal plot of the observed bcc-Fe (110) reflections in a voxel that was heated to temperature above the transition is consistent with the topotactic model.

## The second triple point: fcc-hcp-bcc

The second triple point fcc-hcp-bcc (TP') is determined by intersecting the hcp-fcc and hcp-bcc transition lines. The hcp-fcc and hcp-bcc lines are determined by linear fit to the experimental data points determined in this study. At a given $P$ point, the transition temperature of hcp-fcc is taken to be the average temperature of the lowest temperature XRD patterns that show *in-situ* or quenched fcc-Fe ($T_{fcc}$) and the highest temperature of XRD patterns that show only hcp-Fe ($T_{hcp}$). The lowest $P$-$T$ point for the hcp-fcc transition is anchored at 7.3GPa-820K (*4*). The hcp-bcc transition boundary is determined by $P$-$T$ points based on the texture data in the $T$-quenched XRD (Table S1). The temperature uncertainties for each $P$-$T$ point are estimated as the difference between $T_{fcc}$ and $T_{hcp}$ for the hcp-fcc transition, and $T_{bcc}$ and $T_{hcp}$ for the hcp-bcc transition, added to an instrument uncertainty of ±50K in the spectro-radiometric temperature measurement. The hcp-fcc and hcp-bcc transition lines, with weighted uncertainties, are shown in Fig. S9, overlaying with the 3σ confidence bands. The intersection of the hcp-fcc and hcp-bcc 3σ bands defines the likely location of the TP', constrained at 79±13GPa and 2833±212K. (Fig. S9).

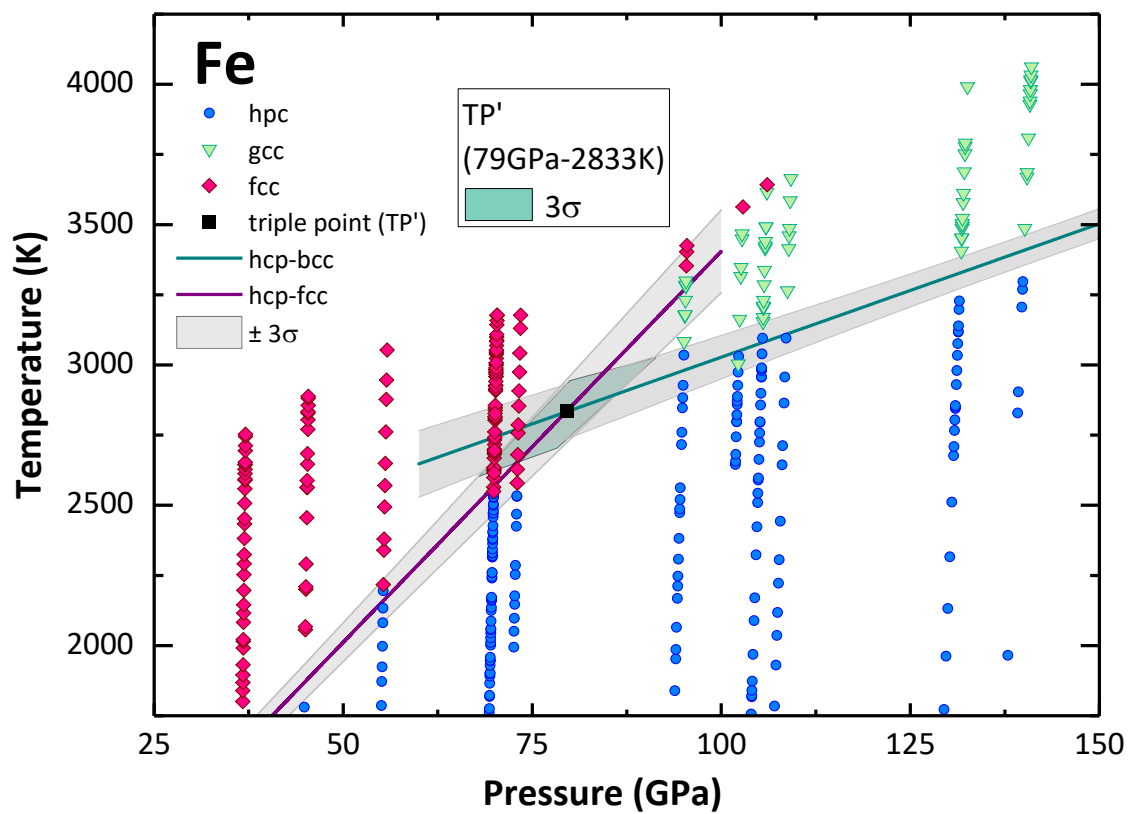

**Fig. S9.** The location of the triple point of fcc-hcp-bcc (TP') determined from the experimental data in this study.

**Table S1. Experimentally obtained hcp-bcc transition data points.**

| P (GPa) | ± P$_{error}$(GPa) | T (K) | ± T$_{error}$(K) |
|---|---|---|---|
| 95 | 2 | 3054 | 79 |
| 101 | 2 | 3017 | 64 |
| 104 | 2 | 3180 | 82 |
| 108 | 3 | 3305 | 135 |
| 131 | 3 | 3391 | 74 |
| 140 | 3 | 3648 | 142 |
| 162 | 4 | 3731 | 98 |
| 169 | 5 | 3874 | 107 |
| 173 | 5 | 4253 | 72 |
| 181 | 5 | 3014 | 103 |
| 186 | 5 | 3669 | 149 |
| 212 | 6 | 3782 | 82 |
| 221 | 6 | 4033 | 104 |